\documentclass[multphys,vecphys]{svmult}

\usepackage{graphicx}        % standard LaTeX graphics tool
\usepackage{multicol}        % used for the two-column index
\usepackage[bottom]{footmisc}% places footnotes at page bottom
\def\mnras{MNRAS}

\def\apj{ApJ}

\def\aj{AJ}
\def\pasp{PASP}
\def\pasj{PASJ}

\begin{document}

\title*{Environmental Effects on Galaxy Evolution Based on the Sloan Digital Sky Survey}
\titlerunning{Environmental Effects on Galaxy Evolution Based on the SDSS} 
\author{Tomotsugu Goto}
\institute{Institute of Space and Astronautical Science,\\ Japan Aerospace Exploration Agency,\\  3-1-1 Yoshinodai, Sagamihara, Kanagawa 229-8510, Japan, 
\texttt{tomo@ir.isas.jaxa.jp}
%\and Name and Address of your Institute \texttt{name@email.address}
}
\maketitle

\begin{abstract}
 We have constructed a large, uniform galaxy cluster catalog from the Sloan Digital Sky Survey data. 
 By studying the  morphology--cluster-centric-radius relation, we have found two
 characteristic environments where galaxy morphologies change
 dramatically, indicating there exist two different physical mechanisms responsible for the cluster galaxy evolution. 
 We also found an unusual population of  galaxies, which have spiral morphology but do not have any emission lines, indicating these spiral galaxies do not have any on-going star formation activity. More interestingly, these  passive spiral galaxies preferentially exist in the cluster outskirts. Therefore, these passive spiral galaxies are likely to be  a key galaxy population being transformed from blue, star-forming galaxies into red, early-type galaxies.
\end{abstract}

%Your text goes here. Separate text sections with the standard \LaTeX\
%sectioning commands.

\section{The Sloan Digital Sky Survey}
\label{goto:sec:1}
% Always give a unique label
% and use \ref{<label>} for cross-references
% and \cite{<label>} for bibliographic references
% use \sectionmark{}
% to alter or adjust the section heading in the running head
%Your text goes here. Use the \LaTeX\ automatism for your citations
 The Sloan Digital Sky Survey (SDSS)\cite{2005AJ....129.1755A} is both an
 imaging and spectroscopic survey of a quarter of the
 sky. Imaging part of the survey takes CCD images of the sky in five
 optical bands ($u,g,r,i$ and $z$)\cite{1995PASP..107..945F}. The
 spectroscopic part of the survey observes one million galaxies. 
 Due to its large quantity and superb quality, the SDSS provides us with an excellent data set to tackle the long standing problems on environmental effects on galaxy evolution.

\section{The SDSS Cut \& Enhance Galaxy Cluster Catalog}

% The imaging part of the SDSS has a potential to produce a cluster
% catalog which can replace the Abell
% galaxy cluster cataglog (Abell 1958, Abell, Corwin and Olowin 1989),
% which has been widely used in astronomical research and produced many
% important results.
 The SDSS Cut \& Enhance galaxy cluster catalog (CE)\cite{2002AJ....123.1807G} is
 one of the initial attempts to produce a cluster catalog from the SDSS
 imaging data. It uses generous color-cuts to eliminate fore- and
 background galaxies when detecting clusters. Its selection function is
 calculated using a Monte Carlo simulation. The accuracy of photometric
 redshift is $\Delta$z=0.015 at z$<$0.3.  For a cluster catalog from the spectroscopic sample of the SDSS, see \cite{2005MNRAS.357..937G}\cite{2005MNRAS.356L...6G}\cite{2005MNRAS.359.1415G}.

\section{The Morphology-Density Relation}
\label{goto:sec:2}

\begin{figure}[t]
\begin{center}
\includegraphics[width=5.8cm,angle=0]{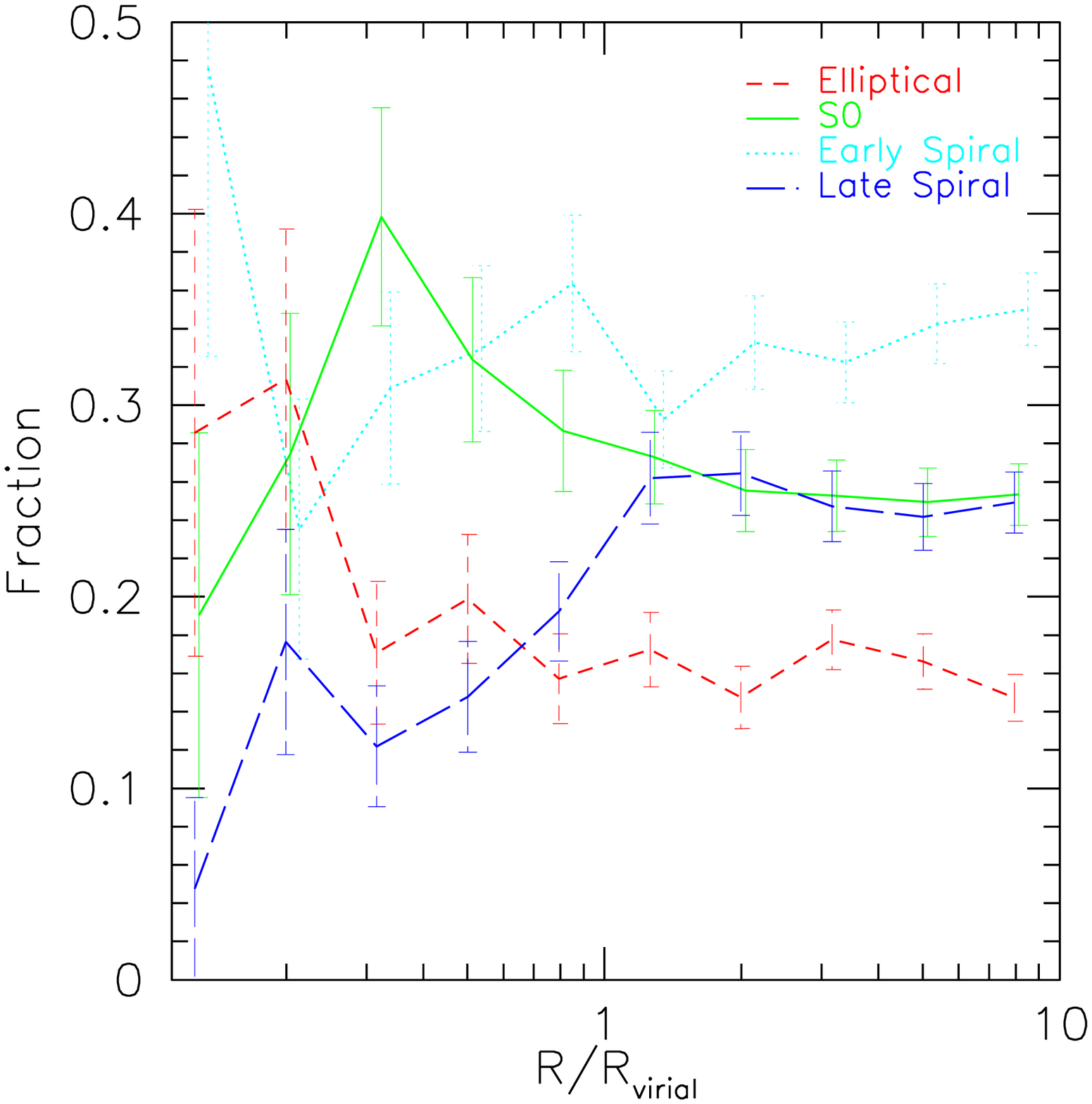}
\includegraphics[width=5.8cm,angle=0]{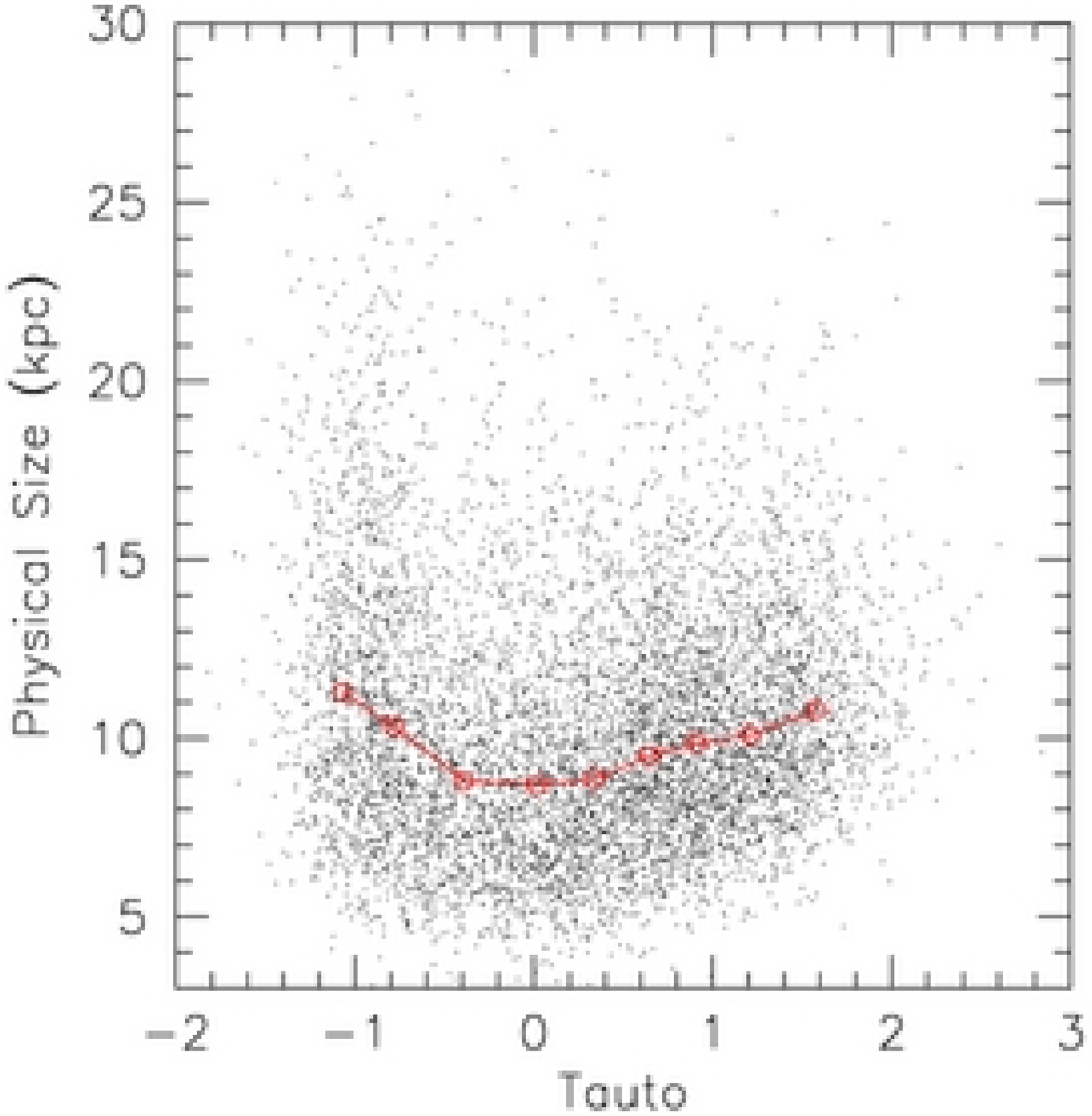}
\end{center}
   \caption{ (left) The morphology-radius relation. Fractions of each type of galaxies are plotted against cluster-centric-radius to the nearest cluster. The short-dashed, solid,
  dotted and long-dashed lines represent 
   elliptical,  S0, early-spiral and late-spiral galaxies classified
   with the automated method ($Tauto$), respectively.}
   \label{goto:fig:mr}
   \caption{ (right)
   Physical sizes of galaxies are plotted against $Tauto$. Petrosian 90\%
   flux radius in $r$ band is used to calculate physical sizes of
   galaxies. A solid line shows medians. It turns over around
   $Tauto\sim$0, corresponding to S0 population.
   }\label{goto:fig:size} 
\end{figure}

 Using a volume limited sample of 7938 spectroscopic galaxies
 (0.05$<$z$<$0.1, $Mr<$-20.5), we investigated the morphology--cluster-centric-radius
 relation in the SDSS \cite{2003MNRAS.346..601G}. We classified galaxies using
 the $Tauto$ method, which utilizes concentration and coarseness of galaxies
 (see \cite{2005AJ....130.1545Y} for more details of $Tauto$). We measured the
 distance to the nearest cluster 
 using the C4 cluster catalog\cite{2005AJ....130..968M}. In Fig.\ref{goto:fig:mr},
 morphological fractions of E, S0, Sa and Sc galaxies are shown in the short-dashed, solid,
 dotted and long-dashed lines as a function of cluster-centric-radius. Around 1
 $R_{vir}$, fractions of Sc start to decrease. Around 0.3 $R_{vir}$, S0
 starts to decrease and E starts to increase. These two changes imply
 there might be two different physical mechanisms responsible for
 cluster galaxy evolution. Since a physical size of S0 galaxies ($Tauto$=0) is
 smaller than E and Sc ($Tauto$=2 and -1) in Fig.\ref{goto:fig:size}, the results are consistent
 with the hypothesis that in the cluster outskirts, stripping creates small S0
 galaxies from spiral galaxies and, in the cluster cores, the
 merging of S0s results in a large Es.

\section{The Passive Spiral Galaxies}
\label{goto:sec:3}

\begin{figure}[t]
\centering{
\includegraphics[width=5.6cm,angle=0]{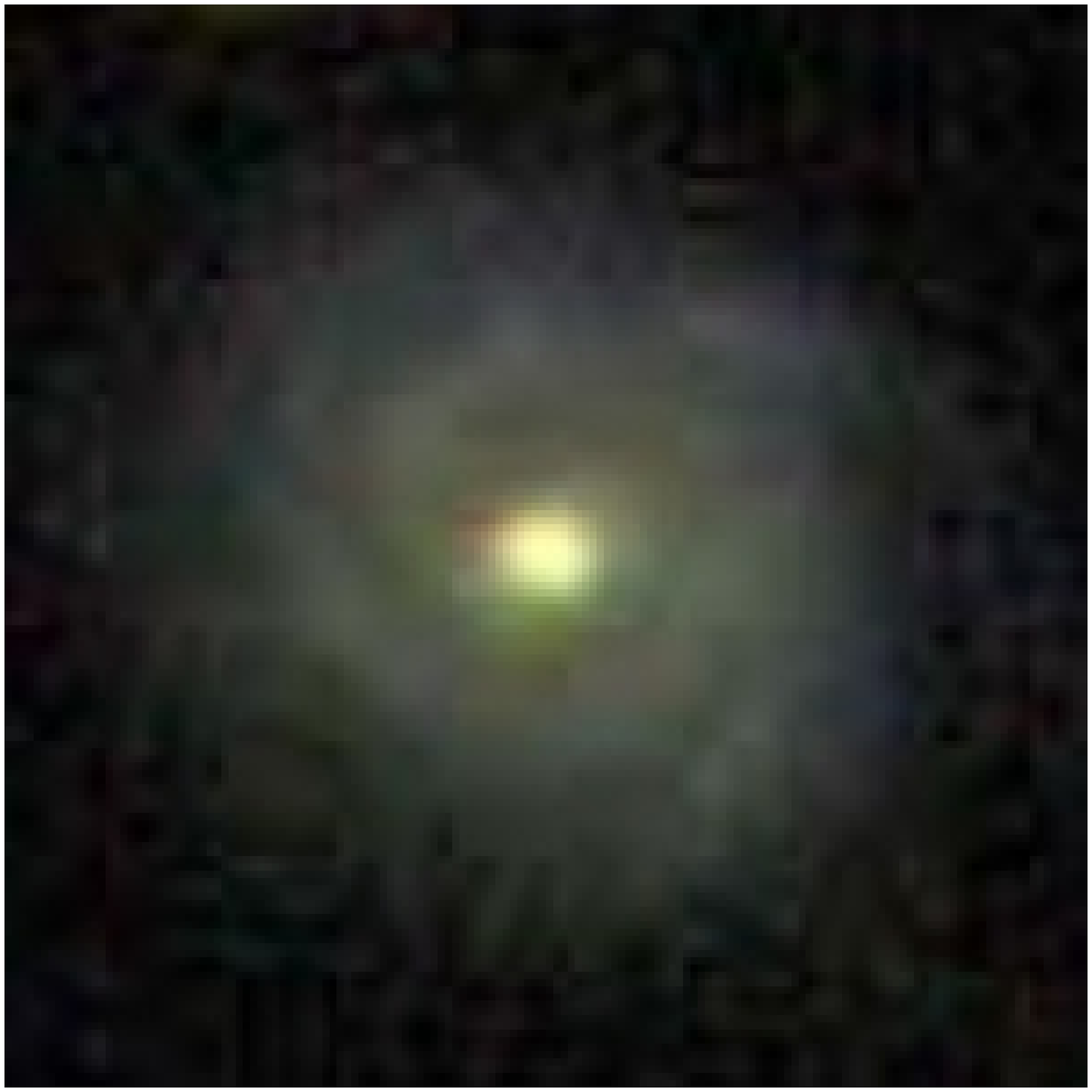}
}
\centering{
\includegraphics[width=5.6cm,angle=0]{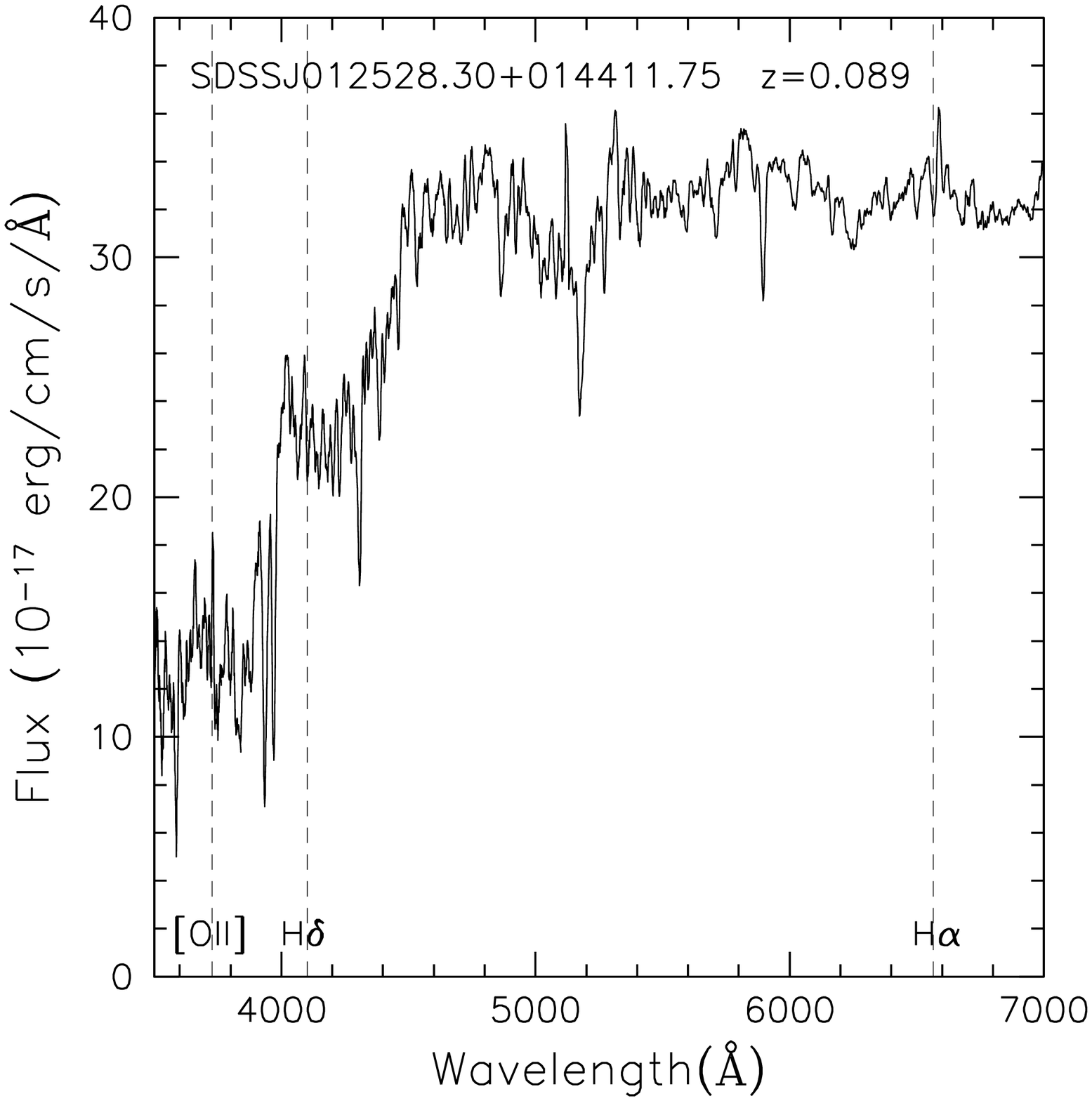} 
}
\caption{ (left)
 An example image of a passive spiral galaxy. The image is a composite of
 SDSS $g,r$ and $i$ bands, showing
 30''$\times$30'' area of the sky with its north up.
 Discs and spiral arm structures are recognized. 
} \label{goto:fig:image}
\caption{ (right)
 An example restframe spectrum of the passive spiral galaxy. The spectrum is
 shifted to restframe and smoothed using a 10\AA\ box.
 The image is shown in Fig. \ref{goto:fig:image}. 
}\label{goto:fig:spectra}\end{figure}

\begin{figure}[t]
\begin{center}
\includegraphics[width=5.8cm,angle=0]{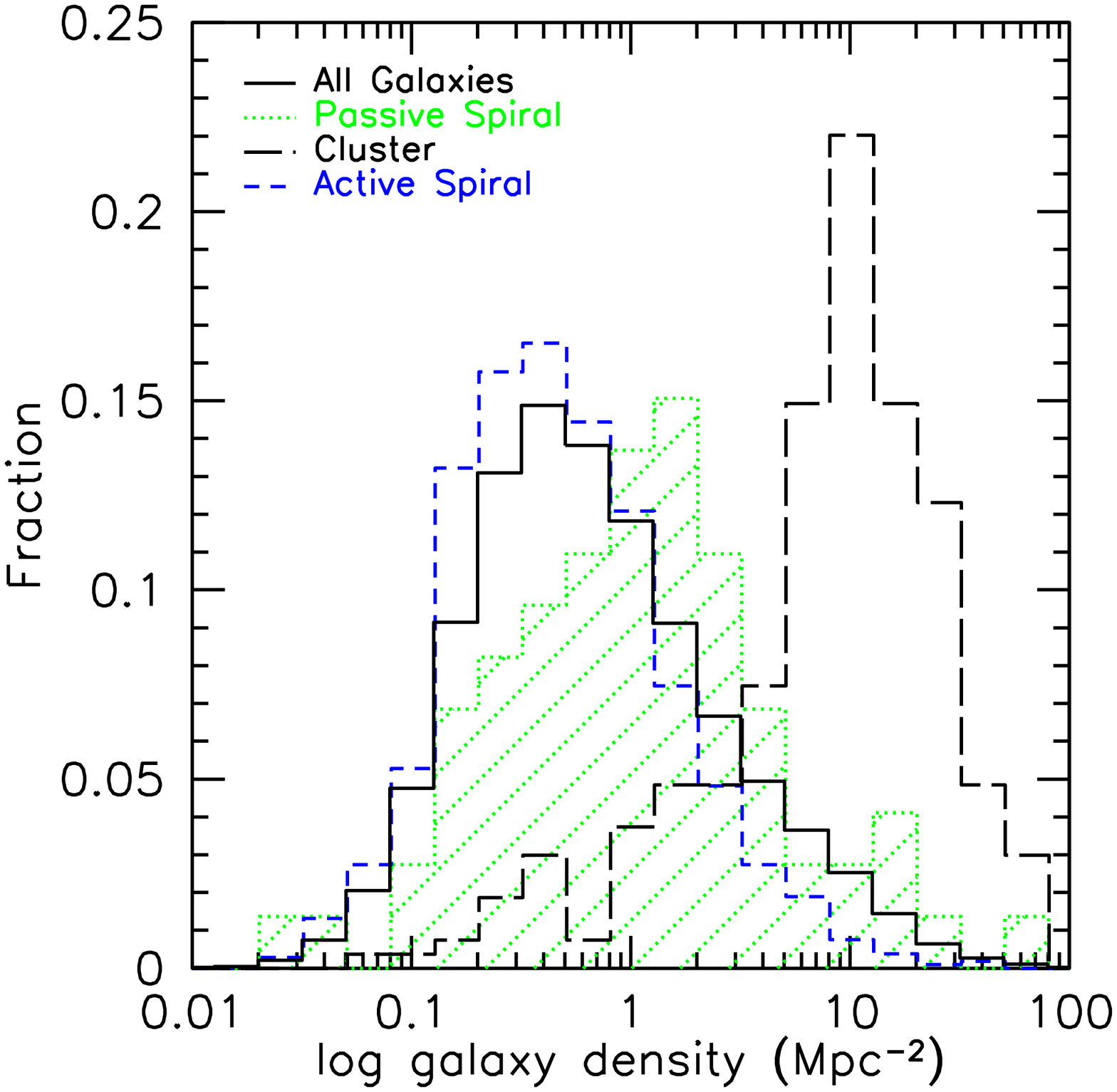}
\includegraphics[width=5.8cm,angle=0]{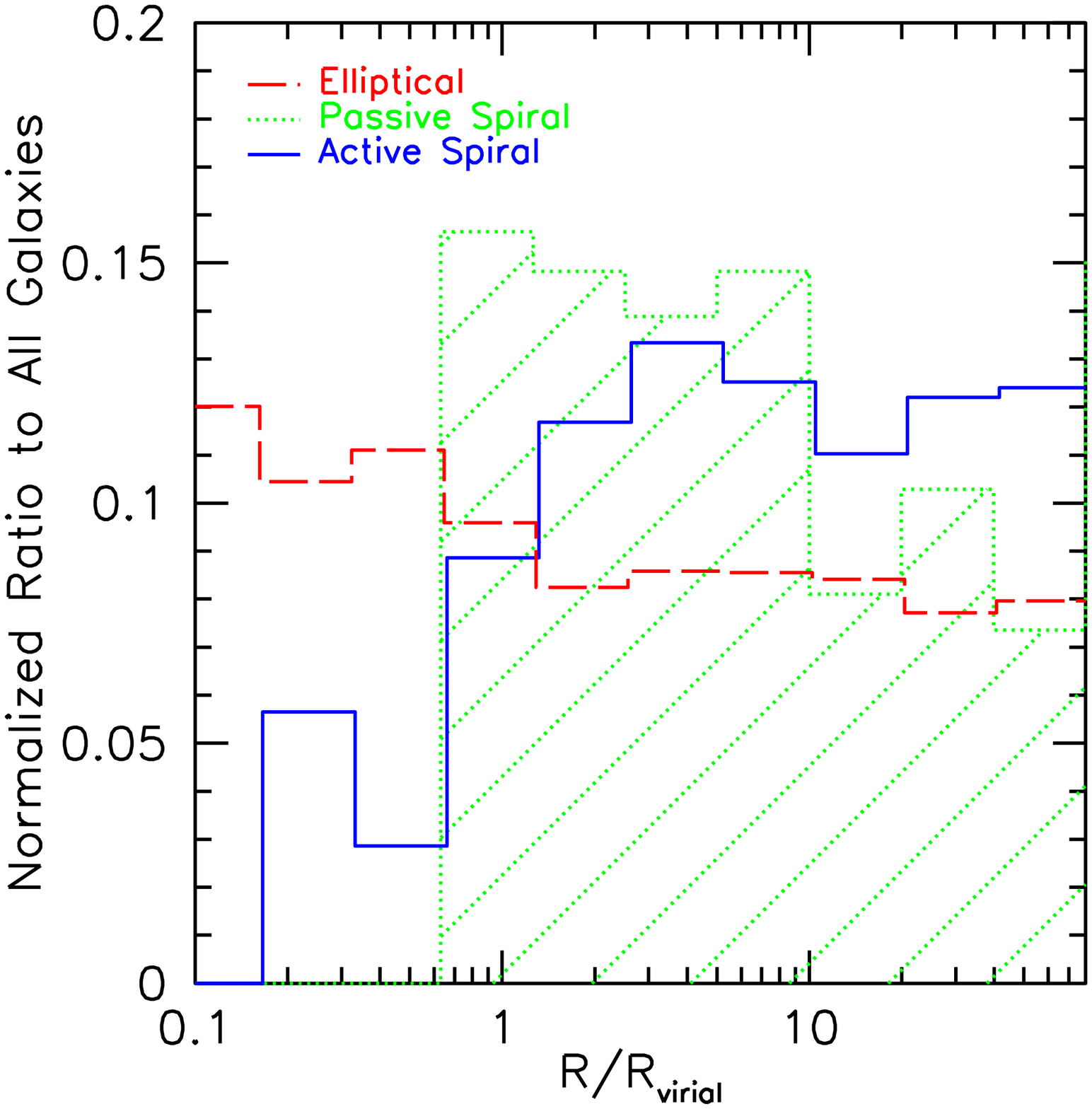}
\end{center}
\caption{(left)
 The density distribution of passive spiral galaxies (hashed
 region) and all galaxies (solid line) in
a volume limited sample. Local galaxy density is measured based on the
 distance to the 5th nearest galaxy within $\pm$3000 km/s.  A Kolomogorov-Smirnov test shows distributions
 of passive spirals and
all galaxies are significantly different. A long dashed line shows
 the distribution of cluster galaxies. A short dashed line shows that of
 active spiral galaxies. Both of them are statistically different from
 that of passive spirals.} \label{goto:fig:density}
\caption{(right)
 The distribution of passive spiral galaxies as a function of
 cluster-centric-radius. The solid, dashed and dotted lines show the
 distributions of passive spiral, elliptical and active spiral galaxies,
 respectively. The distributions are relative to that of all galaxies in
 the volume limited sample and normalized to be unity for clarity.
 The cluster-centric-radius is measured as a distance to a
 nearest C4 cluster\cite{2005AJ....130..968M} within $\pm$3000 km/s, and normalized by
 virial radius \cite{1998ApJ...505...74G}. 
  }\label{goto:fig:radius}
\end{figure}

%Your text goes here.
  In a similar volume limited sample of the SDSS, we have found an interesting class of galaxies with spiral  morphologies (Fig. \ref{goto:fig:image}), and without any star formation activity (shown by the lack of emission lines in the spectrum; Fig. \ref{goto:fig:spectra}). 
% In addition, we have found  an unusual 
% population of galaxies, which have spiral morphologies, but do not show
% any star-formation activity, indicated by the lack of any emission lines. 
 These galaxies are called ``passive spiral galaxies'', and  interesting since 
 they are against currently favored galaxy formation models in which star-formation is sustained by the density waves.
  Using a volume-limited sample (0.05$<z<$0.1 and $Mr^*<-20.5$; 25813
 galaxies) of the  SDSS data, we found 73 (0.28$\pm$0.03\%)
 passive spiral galaxies and studied their environments.
 It is found that passive
 spiral galaxies exist in a local galaxy density of 1--2 Mpc$^{-2}$ and
 have a 1--10 cluster-centric virial radius (Figs. \ref{goto:fig:density} and \ref{goto:fig:radius}; \cite{2003PASJ...55..757G}). In other words, passive spirals preferentially exist in the cluster infalling regions.  This is the first direct evidence to connect the origin of passive spiral galaxies to cluster related physical mechanisms. 
 These characteristic environments
 coincide with a previously reported environment where the galaxy
 star-formation rate suddenly declines\cite{2004AJ....128.2677T} and the so-called 
 morphology-density relation turns (See Section \ref{goto:sec:2}).
 It is likely that the same physical mechanism is
 responsible for all of these observational results in the cluster infalling regions.
  The existence of passive spiral
 galaxies suggests that a physical mechanism that works calmly (e.g., gas stripping) is
 preferred to dynamical origins such as major merger/interaction since such a
 mechanism would destroy the spiral-arm structures. 
 Passive  spiral galaxies are likely to be a key galaxy population in
 transition between red, early-type galaxies in low-redshift clusters
 and blue, spiral galaxies more numerous in higher-redshift clusters.

%More interestingly, these galaxies
% exist in the infalling region of clusters (1$\sim$10 R$_{vir}$ or
% 1$\sim$2 Mpc$^{-2}$; Fig.\ref{fig:density},
% Fig.\ref{fig:radius}). These passive spiral galaxies might be the
% population of galaxies in transition between blue/spiral and red/S0
% galaxies. Details are given in Goto et al. (2003c)

\section{Conclusions}
 We have  constructed a large, uniform galaxy cluster catalog from the SDSS
 data in order to investigate physical mechanisms responsible for the cluster galaxy evolution.
By revealing the morphology--cluster-centric-radius relation with the largest statistics to date, we have found two characteristic environments where galaxy morphology changes
 dramatically, suggesting the existence of two different physical
 mechanisms in the cluster regions. In addition, we have found unusual galaxies with  spiral morphology but without any emission lines. More interestingly, we found that these passive spiral galaxies preferentially exist in the cluster infalling regions, indicating that they are likely to be transition objects between field star-forming spirals and red, cluster early-type galaxies due to the environmental effects in the cluster outskirts.

%\begin{equation}
%\vec{a}\times\vec{b}=\vec{c}
%\end{equation}

%\subsubsection{Subsubsection Heading}
%Your text goes here. Use the \LaTeX\ automatism for cross-references as
%well as for your citations, see Sect.~\ref{sec:1}.
%
%\paragraph{Paragraph Heading} %
%Your text goes here.
%
%\subparagraph{Subparagraph Heading.} Your text goes here.%
%
\index{paragraph}
\end{document}